# Coherent control of a strongly driven silicon vacancy optical transition in diamond

Yu Zhou[1], Abdullah Rasmita[1], Ke Li[1], Qihua Xiong[1], Igor Aharonovich[2,3], Wei-bo Gao[1]

[1] *Division of Physics and Applied Physics, School of Physical and Mathematical Sciences, Nanyang Technological University, Singapore 637371, Singapore*

[2] *School of Mathematical and Physical Sciences, University of Technology Sydney, Ultimo, NSW, 2007, Australia*

3. *Institute of Biomedical Materials and Devices (IBMD), Faculty of Science, University of Technology Sydney, Ultimo, NSW, 2007, Australia*

**The ability to prepare, optically read out and coherently control single quantum states is a key requirement for quantum information processing. Optically active solid state emitters have emerged as promising candidates with their prospects for on chip integration as quantum nodes and sources of coherent photons for connecting these nodes. Under strongly driving resonant laser field, such quantum emitter can exhibit quantum behavior such as Autler-Townes splitting and Mollow triplet spectrum. Here we demonstrate coherent control of a strongly driven optical transition in silicon vacancy (SiV) center in diamond. Rapid optical detection of photons enabled the observation of time resolved coherent Rabi oscillations and the Mollow triplet from an optical transition of a single SiV defect. Detection with a probing transition further confirmed Autler-Townes splitting generated by a strong laser field. Coherence time of the emitted photons is shown to be comparable to its lifetime and robust under very strong driving laser field, which is promising for generation of indistinguishable photons.**

## Introduction

Coherent control of atom-photon interfaces is vital in the realization of quantum information protocols. In particular, the interface between solid state qubits and photons serves as promising candidates for practical and scalable quantum technologies [1-5]. Different types of defects in solids have been studied so far, and many of them show excellent optical or spin properties to be proper quantum qubits candidates. Among others, the nitrogen-vacancy (NV) centers in diamond have achieved partial success in spin-photon interface and spin-spin entanglement mediated by photons[1-2]. However, despite the long spin coherence times of the NV, the defect suffers from low percentage (3-5%) of the total emission into its weak zero phonon line (ZPL) and from strong inhomogeneous broadening. Recently, increased efforts have been put in place to identify other solid state qubits with improved inherent properties. The silicon vacancy (SiV) center in diamond is one potential candidate, with ~ 70% of the photons emitted into the ZPL[6-7]. Different SiV defects exhibit intrinsically identical spectral properties in low-strain bulk diamond[8] and nearly transform-limited linewidth[9-10]. SiV spin coherence times are at the range of ~ 40 ns at cryogenic (4K) temperatures[11-12]. Important to note, the coherence times are only limited by spin relaxation time and are estimated to increase dramatically at lower temperatures[13]. As a result, SiV defect has become a promising candidate to be a key building block for quantum information processing.

Towards this goal, preparation and coherent control of the emitted photons from SiV is a prerequisite. For a two-level system under strong pumping resonant field excitation, state population will oscillate between the ground and the excited states, also known as Rabi oscillations. In the frequency domain, the emission under continuous wave (cw) laser excitation will exhibit Mollow triplet spectrum[17], which is a hallmark for quantum coherent control and enables a robust approach to generate single photons with detuned frequency from the resonance[18-26]. In this work, we demonstrate coherent control of a strongly driven single SiV defect in diamond. In particular, we observe Rabi oscillation with fast photon detection using both a nanosecond laser pulse and a cw laser field. Fourier transform of the time-resolved detection combined with a laser detuning, results in an observation of the Mollow triplet spectrum. In the frequency domain, the Autler splitting has been confirmed by a Λ energy scheme[11-12]. In the above measurements, photon coherence, which is critical for photon interference in building a quantum network, has been characterized in both low and high power regime.

The SiV defect consists of an interstitial silicon atom, neighboring two vacancies along a [111] crystallographic axis in diamond[27], as shown in figure 1a. The defect is negatively charged and has a $D_{3d}$ symmetry which is inert to strain and fluctuating electric fields[14,15,28]. The electronic structure and optical transitions of the negatively charged SiV in diamond have been characterized in details recently[14-16]. Both ground and excited states of the SiV ZPL are split due to spin orbit coupling, resulting in four lines at cryogenic temperature as shown in figure 1b (black solid lines). In our experiment, the transitions are successfully identified by scanning resonant laser and recording the photoluminescence (PL) at the same time (see figure S1). For the current work, the coherent control is performed on transition C as marked in Figure 1b. Transition C is associated with the lower energy levels in both the excited state and ground state in SiV. It doesn't suffer from fast relaxation from the upper branches of the energy levels, therefore offering a better count rate and longer coherence times compared to A, B, D transitions. By resonantly exciting the transition C, and collecting the photons at the phonon sideband above 750 nm, a narrow linewidth of 219 MHz at saturation power is achieved (figure 1c). The linewidth without power broadening is 154 MHz, close to the transform limited linewidth of 86 MHz, calculated from a direct lifetime measurement that is shown in figure 1d.

**Time-resolved Rabi oscillation**

We first demonstrate real-time observation of the time resolved Rabi oscillation measured with a superconducting detector with a short jitter time. In the time domain, a strong resonant cw laser coupling to a two-level system yields oscillation of the excited state population, and can be described by $\sin^2(\Omega_g t/2)$, where $\Omega_g = \sqrt{\Delta^2 + \Omega^2}$ and $\Delta$ is the laser detuning, $\Omega = \mu E/\hbar$ is the bare Rabi frequency, $\mu$ is the optical transition dipole moment and $E$ is the driving field amplitude. In the frequency domain, the physics model can be best demonstrated with a dressed state as shown in figure 1b (red dashed lines). Under a strong laser field, the ground state and excited state will split (Autler-Townes splitting) into two states separated by an energy $\Omega_g = \sqrt{\Delta^2 + \Omega^2}$. Two frequencies of the emission from four transitions are degenerate and therefore the dressed state will result in an emission pattern with three Lorentzian lines with center frequency $v_0 + \Delta$ and $v_0 + \Delta \pm \Omega_g$.

In our experiment, a 5 ns laser pulse is used to excite a single SiV center. The emitted photons are collected and analyzed with a time-correlated single photon counting system (see Methods). As shown in figure 2a, the Rabi oscillations are clearly observed with a shorter oscillation period corresponding to a higher excitation power. The curves are fit with the theoretical value of excited state population which can be expressed as (up to a displacement time $dt$) [20]

$$P = 1 - e^{-\eta|\tau|}\left(\cos(\mu|\tau|) + \frac{\eta}{\mu}\sin(\mu|\tau|)\right) \quad (1)$$

Here $\eta = 1/2T_1 + 1/2T_2$ and $\mu = \sqrt{\Omega_g^2 + (1/2T_1 - 1/2T_2)^2}$ with $T_1$ and $T_2$ are the lifetime and the photon coherence time, respectively. In the fitting, $T_2$ and $\Omega_g$ are used as free parameters while the measured $T_1$ value is $1.85 \pm 0.02 ns$. Extracted Rabi frequency is linearly proportional to the square root of the excitation power, confirming the Rabi oscillation behavior (figure 2b). Average value of $T_2$ is calculated to be 1.62 ns, smaller than the ideal value $2T_1 = 3.7 ns$. This is likely due to the pure dephasing of the excited state. More importantly, as shown in figure 2c, the value of $T_2$ stays above 1.3 ns even with Rabi frequency of 12 GHz corresponding to 300 times of saturation power. This shows that no apparent excitation induced dephasing is observed. This can be partially explained by the weak electron-phonon coupling of the SiV, and the reduced sensitivity to fluctuating electric fields within the diamond lattice[8-10]. Robust $T_2$, combined with the frequency stability and efficient generation of SiV single photons shows the feasibility of SiV as an efficient and coherent single photon source.

The plot of time-dependent fluorescence intensity profile as a function of laser detuning is shown in figure 2d. With detuned laser, the oscillation frequency is increased to the generalized Rabi frequency $\sqrt{\Omega^2 + \Delta^2}$. To best illustrate this phenomenon, a Fourier transform of each curve is taken and the frequency components are shown as a function of the laser detuning. Three peaks corresponding to the three frequency components of the Mollow triplet can be clearly seen. The dashed lines in the figure 2e corresponds to the theoretical center frequency $v_0 + \Delta \pm \sqrt{\Omega^2 + \Delta^2}$ with $\Omega$ extracted when the laser detuning is 0.

## Photon correlation measurement

To study the quantum nature of a single SiV defect in more details, the second order correlation function, $g^{(2)}(\tau) = \langle I(t)I(t+\tau)\rangle / \langle I(t)\rangle^2$, is recorded as a function of excitation power using a Hanbury-Brown-Twiss (HBT) setup. For a perfect single photon emitter, $g^{(2)}(0) = 0$. In our experiment (figure 3a), the resonantly excited SiV shows a $g^{(2)}(0) = 0.061 \pm 0.026$ at saturation power, without correcting for the detector response time. When the transition is driven strongly (above saturation), oscillations with different periods are observed in the delay time range of $[0, \pm 5]ns$. Unlike single photon detection experiment in the above paragraphs where the time 0 is set by the excitation pulse rising time, here the start signal is set by the detection of first emitted photon in the $g^{(2)}(\tau)$ measurement. The first photon (start signal) prepares the system into the ground state, and the second photon (stop signal) will show the population of the excited state, which varies due to the Rabi oscillations. The measured $g^{(2)}(\tau)$ data (figure 3a) is fit with the same equation (1) up to a scaling factor, A, and displacement count $\Delta g$ due to the effect of background and finite detector's impulse response function. A single set of $T_1 = 1.85 \pm 0.02ns$ and $T_2 = 1.62ns$ fits the data well at different excitation powers. The Rabi frequencies extracted from the fitting are plotted as a function of the square root of the laser power, showing an expected linear relationship (figure 3b) comparable to the one obtained from time resolved Rabi oscillations experiment.

## Autler-Townes splitting

In addition to the Fourier transform analysis of the time-resolved Rabi oscillations, the Autler-splitting can be observed in the frequency domain. To demonstrate this, the same transition C is resonantly strongly driven, and is probed using transition D. The energy level structure to probe the splitting is shown in figure 4a. Consider a $\Lambda$ system which includes a strongly pumped transition C ($\Omega_C^2$ much larger than $P_{sat}$) and a weakly probed transition D. C and D transitions have different ground states and share the same excited state. A pump laser is tuned to C transition and a probe laser is scanned across D transition. The detuning of the laser relative to the C (D) transition is indicated by $\Delta_C(\Delta_D)$ and the Rabi frequencies of the C (D) transitions are denoted as $\Omega_C(\Omega_D)$. Similarly to the coherent population trapping, the fluorescence intensity will show a dip around $\Delta_D = 0$ due to the destructive interference between the dipole moment of transition C and D. If one of the transitions is strongly pumped, the splitting related to the observed dip will correspond to the splitting of the excited state also known as Autler-Townes splitting.

In the experiment we vary the laser power for the C transition while power of the probe laser for the D transition is fixed. The normalized fluorescence intensity as a function of $\Delta_D$ is shown in figure 4b. The value of $\Omega_C$ is obtained by fitting the data with the Lindblad master equation (details can be found in the Supplementary). The value of the dip width is larger, as the power pumping on transition C increases, consistent with the Autler-Townes splitting. The $\Omega_C$ values extracted from the fitting are plotted as a function of the square root of the laser power, showing an expected linear relationship (figure 4c) comparable to the one obtained from time resolved Rabi oscillation experiment. To get more information on the fluorescence characteristics as a function of the laser detuning, intensity measurements are recorded by varying $\Delta_C$ and $\Delta_D$ while keeping the C transition excitation power fixed at ~ $20 P_{sat}$. The measured data is shown in figure 4d, where the yellow dotted line is guide to the eye and the data is in a good agreement with the simulated value (figure 4e).

## Rabi oscillation and Ramsey fringe with pulse excitation

In the last part, we demonstrate coherent control of SiV optical transition with a pulsed laser. To avoid cross-excitation of other unwanted transitions, a 200 picosecond laser with 80MHz repetition rate is generated after a two-stage filtering setup based on spectrometer and also a fused silica etalon (Light machinery, FSR 30GHz, Finesse 20, See supplementary information Figure.S3. for detailed setup). As shown in the Figure 5a, the photon counts show Rabi oscillations with the increase of the laser power coupled to the optical transition. As expected, the oscillation followes a Sine curve as a function of the square root of the excitation power. Furthermore, two π/2 pulses with a delay τ are used to measure the Ramsey fringes. The photon counts oscillate as the delay changes by a length of wavelength. As shown in Figure 5b, we extract the maximum and minimum of the Ramsey fringe envelopes around delay τ. The visibility of the oscillation as a function of the delay τ is shown in Figure 5c. With τ lager, the visibility will be smaller which is related to both fluorescence lifetime and dephasing rate $T_2^*$. An exponential decay fitting of the visibility of Ramsey yields $T_2$ =780±141 ps, which is consistent with the linewidth measurement for this transition.

## Conclusions

In summary, we demonstrated coherent control of a strongly driven optical transition of a single SiV defect in diamond. SiV defects studied in this work are embedded in nanodiamond particles. Unlike the NV center that only exhibits excellent quantum properties when embedded in a large ultra-pure bulk crystal, the availability of suitable quantum systems in nanodiamonds offers a great advantage. For instance, it opens up possibilities of engineering scalable hybrid quantum photonic systems by manipulating the nanodiamond hosting the SiV defect onto a photonic crystal cavity or a waveguide to engineer quantum registers. Coherence time of SiV emission is long and robust even at high excitation powers. This aspect is important for connecting quantum nodes through Hong-Ou-Mandel interference experiments[29]. Several immediate research directions come out in the next steps: The Mollow Triplet photons in each sideband can be separated by the combination of Michelson interferometer and Fabry-Perot filters (FP) [30]. The role of the FB etalons is to filter out the excitation laser and produce frequency tunable single photon source by detuning the excitation laser frequency. A detailed study of the photon statistical signature can be made for the triplet emission[30]. In addition, the coupling of a Mollow triplet sideband to an optical cavity may lead to a strongly enhanced resonant sideband[31-33], enabling tunable dressed state laser[34]. Finally, under magnetic field the spin resolved Mollow triplet may assist in single-shot spin readout[26] or spin-photon interface.

## Methods

**Experimental setup** The sample was mounted on a 4 K cryostation (Montana Instruments). A home-built confocal microscope system was used to excite the sample and collect the fluorescence through a 0.8 NA Nikon objective. In photoluminescence excitation experiment, Titanium: Sapphire (Ti:SAP) laser with less than 5 MHz linewidth was used to excite the sample. Phonon sideband (more than 750 nm) of the emission was collected and guided to an avalanche photo diode through a single mode fiber. In the photon correlation antibunching measurement, the emission was guided to a beam splitter. Two avalanche photo diodes were used to record the counts and connected to a time-correlated photon counting card (Picoharp, PH300). In the time resolved Rabi oscillation experiment, arbitrary waveform generator (Technotronix AWG7000) was used to generate the pulse triggering an electro-optic modulator, and synchronize the whole experiment. The emission was guided to a low jitter ( $\sim 30 ps$ ) super-conducting-single-photon detectors (SSPD). For the Autler-townes splitting detection, two laser (Titanium:Sapphire laser and a diode laser) were combined together using a beam splitter and then guided to the excitation arm. All lasers were locked to a wavemeter with 10 MHz accuracy and feedback speed 500Hz to avoid center wavelength drift. In time-resolved Rabi oscillation, photon correlation and Autler-Townes experiment, we collect emission with polarization perpendicular to the input laser to suppress residual counts from the excitation laser.

**Diamond growth** The diamond sample was grown using a microwave plasma chemical vapor deposition technique. Detonation nanodiamonds (4-6) nm were dispersed on a silicon substrate and used as seeds for the diamond growth. The growth condition was 60 Torr, 950 Watt. The final size of the nanodiamonds was several hundreds of nanometers.

**Note added**: In preparation of this manuscript, we note a paper on Arxiv reporting a complementary method for measuring Rabi oscillation[35].

**Acknowledgments** We thank the discussion with Christoph Becher. We acknowledges the support from the Singapore National Research Foundation through a Singapore 2015 NRF fellowship grant (NRF-NRFF2015-03) and its Competitive Research Program (CRP Award No. NRF-CRP14-2014-02), and a start-up grant (M4081441) from Nanyang Technological University. I. A. is the recipient of an Australian Research Council Discovery Early Career Research Award (Project Number DE130100592). Partial funding for this research was provided by the Air Force Office of Scientific Research, United States Air Force (grant FA2386-15-1-4044). Q.X. gratefully thanks Singapore National Research Foundation via Investigatorship award (NRF-NRFI2015-03) and Ministry of Education via an AcRF Tier2 grant (MOE2012-T2-2-086).

**Author Contributions** All authors contributed extensively to this work.

**Author Information** The authors declare that they have no competing financial interests. Correspondence and requests for materials should be addressed to Igor.Aharonovich@uts.edu.au or wbgao@ntu.edu.sg.

## Figures and Figure Legends

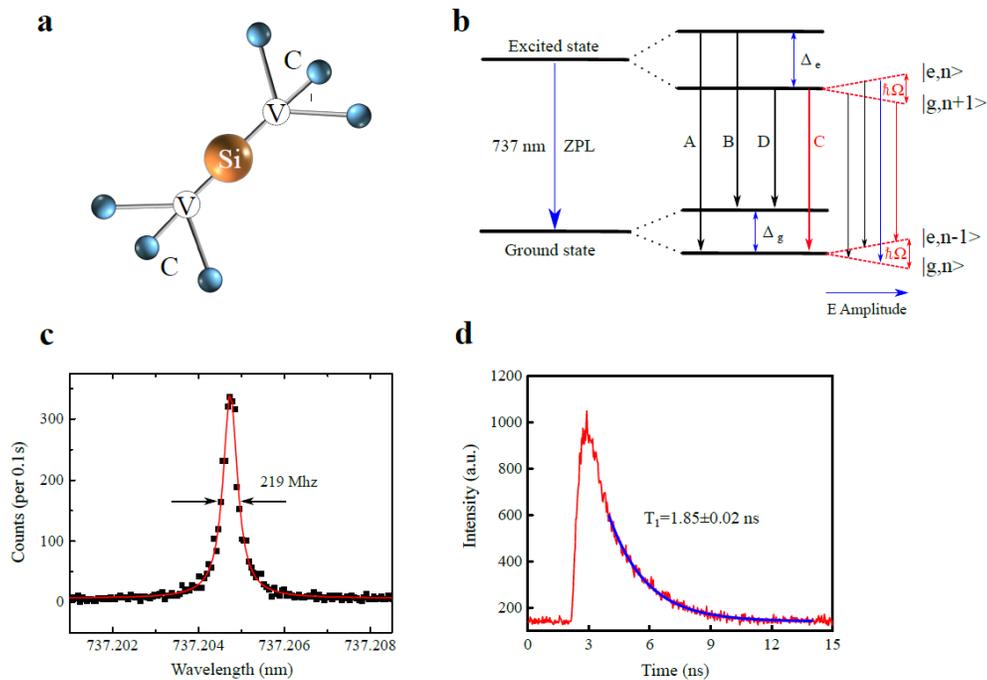

**Figure. 1: The SiV defect. a**. Atomic structure of a silicon vacancy colour canter in diamond. Silicon atom (yellow) is neighbouring two carbon vacancies (labelled V) in a diamond lattice. **b**. Mollow triplet and SiV energy level scheme. The zero phonon line is at 737 nm. The excited and ground states split by an energy $\Delta_e = 410 GHz$ and $\Delta_g = 228 GHz$, respectively, resulting in four transitions labelled as A, B, C, D. With increasing electric field coupling to C transition with a resonant laser, four dressed states are formed and labelled as $|g,n\rangle, |e,n-1\rangle, |g,n+1\rangle$ and $|e,n\rangle$. **c**. Resonant measurement of the C transition linewidth at saturation power. **d**. Lifetime measurement. A resonant laser pulse of ~100 ps is used to excite C transition, resulting in an exponential decay with a lifetime of 1.85ns.

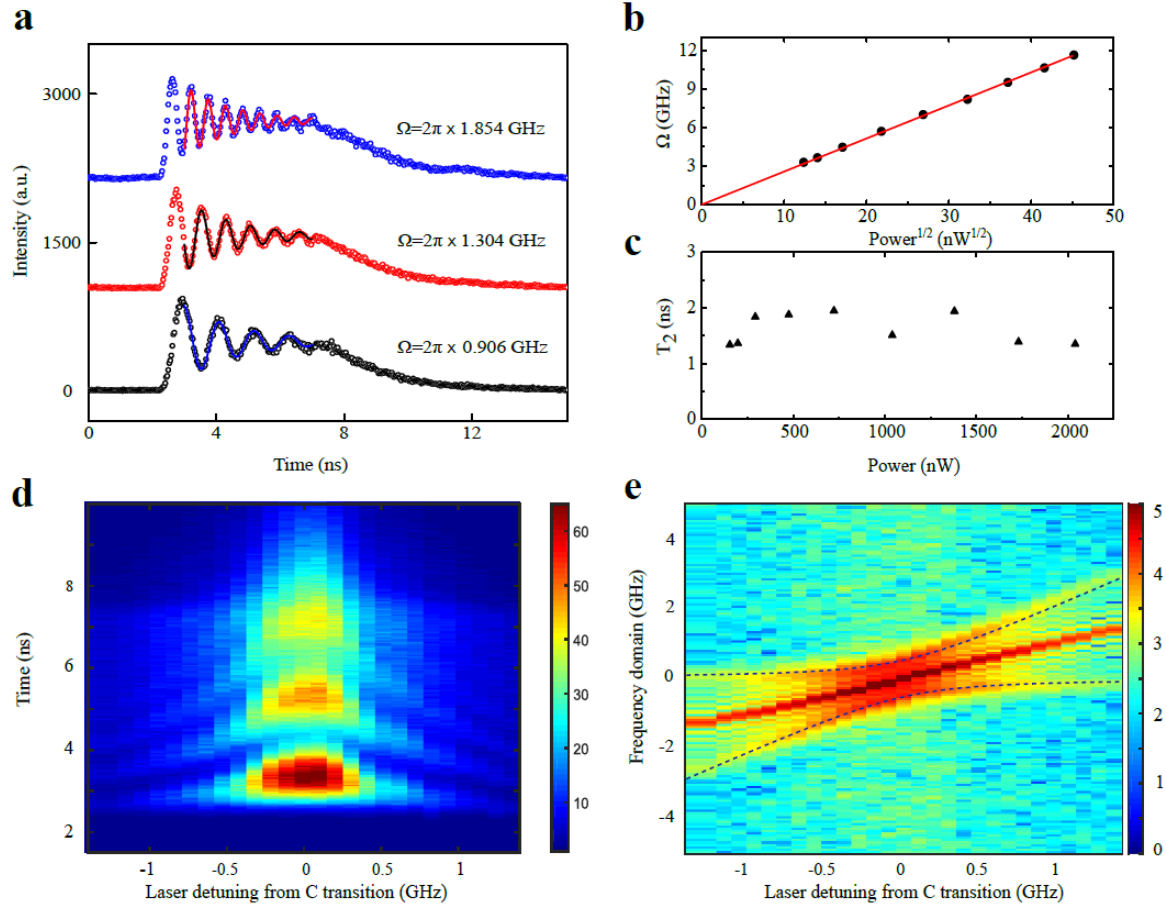

**Figure. 2: Time resolved Rabi oscillation. a**. Rabi oscillations of C transition when excited with a 5 ns laser pulse at three different excitation powers. Laser pulse period is 15 ns. Fitting of the curves with equation (1) results in the Rabi frequency $\Omega/2\pi = 0.906 GHz, 1.304 GHz, 1.854 GHz$. **b**. Fitted Rabi frequency for curves in **a** as a function of the square root of the excitation laser power, exhibiting a linear trend (red line). **c**. The extracted $T_2$ times from the fitting as a function of the excitation laser power. No apparent decrease is seen up to 2000nW, which corresponds to 300 times of a saturation power. **d**. Excitation with a fixed Rabi frequency $\Omega = 2\pi \times 1.304 GHz$ while modifying the laser detuning from C transition. Different colours represent different intensity amplitude. **e**. Fourier Transform analysis of the measurement in **d**. Two side branches fit well with the theoretical values $v_0 + \Delta \pm \sqrt{\Omega^2 + \Delta^2}$.

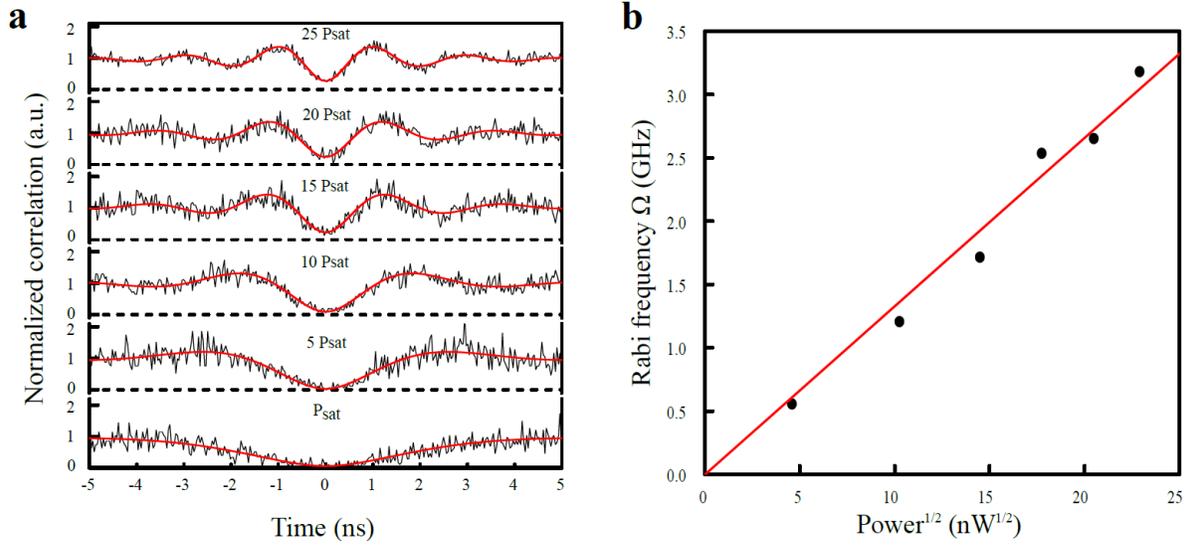

**Figure. 3: Photon correlation measurements. a**. Raw data of the second order correlation function $g^{(2)}(\tau)$, fit with equation (1). The $g^{(2)}(\tau)$ is normalized to 1 at infinite delay times. $P_{sat}$ is the saturation power ~ $20nW$. $T_1 = 1.85 \pm 0.02ns$ and $T_2 = 1.62ns$ are used while $\Omega$ is fitted. **b**. The Rabi frequency as extracted from the fitting is plotted as a function of the square root of the excitation laser power. The values follow a linear trend.

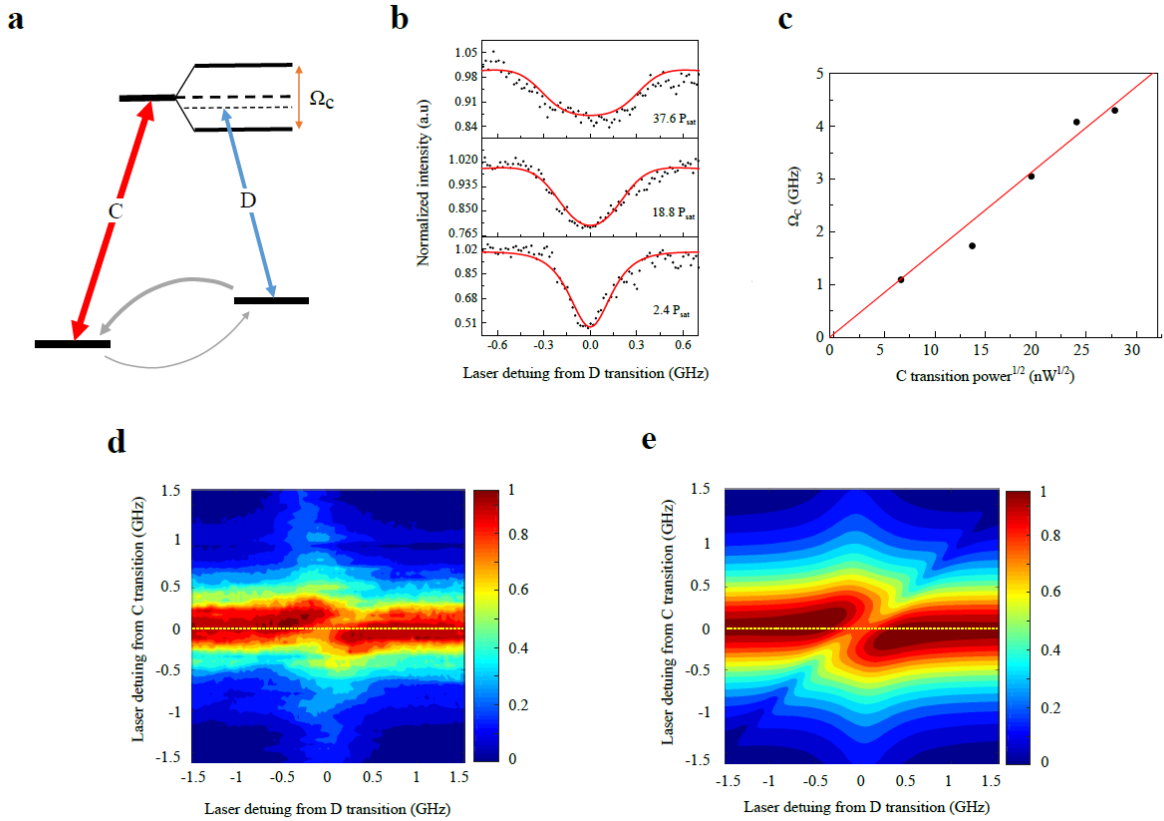

**Figure. 4: Autler-Townes splitting detection.** **a**. Level structure of the Autler-Townes splitting detection scheme. A much stronger laser is pumping resonantly the C transition while a probing laser is detuned on D transition to detect the excited energy levels. Grey solid curves represent the decay or relaxation from each level. **b**. Experimental and simulated curves for three different laser powers exciting the C transition. The frequency of the excitation laser is fixed, and a second laser scans D transition in the range of $[-0.7 GHz, \ 0.7 GHz]$ from resonance. **c**. $\Omega_C$ is extracted from the fit, and plotted as a function of the square root of power of the excitation laser. **d**. Experimental data for measuring the Autler-Townes splitting. SiV fluorescence counts are detected as a function of scanning excitation laser across D transition in the range of $[-1.5 GHz, \ 1.5 GHz]$ from resonance, while excitation laser on C is fixed and stepped in the range of $[-1.5 GHz, \ 1.5 GHz]$ from resonance too. Laser powers are $20 P_{sat}$ and $2.5 P_{sat}$ for C and D transition respectively. **e**. Simulated SiV fluorescence for the experiment in **d** (For simulation method and values, see supplementary).

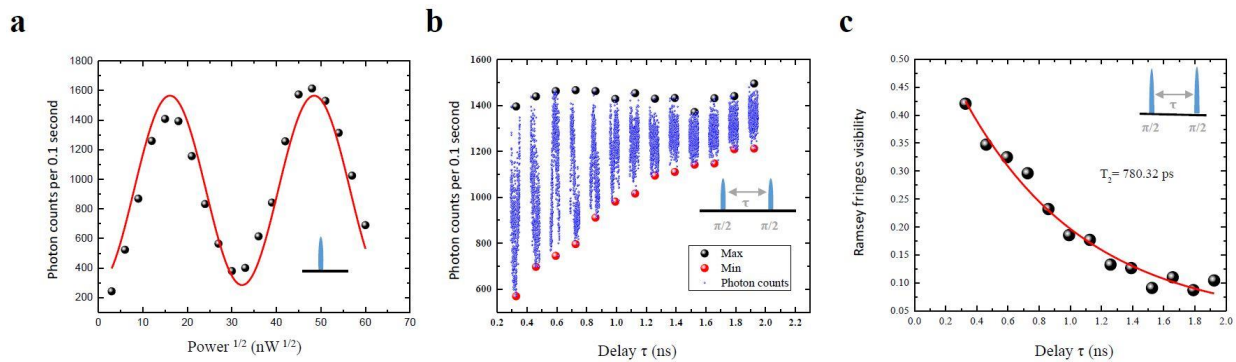

**Figure. 5: Ultrafast control of SiV optical transition.** a. Rabi oscillation with pulse excitation. The photon counts are recorded as function of square root of power, the curve is fitted using a Sine function. b. Photon count oscillates when the delay between two π/2 picosecond pulses varies around τ. Maximum and minimum value of the Ramsey oscillation can be extracted as shown with black and red dots. Blue curves represent the oscillation of the photon counts around delay τ with time shown stretched here for illustration. c. Ramsey fringe visibility as a function of delay τ. An exponential decay fitting yields T2= 780±141 ps.